\documentclass{article}

\usepackage[english]{babel}

\usepackage[letterpaper,top=2cm,bottom=2cm,left=1.3cm,right=1.3cm,marginparwidth=1.5cm]{geometry} %

\usepackage{caption}
\usepackage{subcaption}
\usepackage{bbm}
\usepackage{amsmath, amssymb, mathtools}
\usepackage{graphicx}
\usepackage{amsthm}

\usepackage[colorlinks=true, allcolors=blue]{hyperref}
\usepackage{bm}
\newcommand*{\R}{\mathbb{R}}

\usepackage[numbers]{natbib}
 \usepackage{authblk}
 \usepackage{mathptmx}
\usepackage{etoolbox}
\usepackage{xcolor}%
\usepackage{layout}
\usepackage{multirow}
\usepackage{dcolumn}

\usepackage[utf8]{inputenc}

\title{Co-evolving networks for opinion and social dynamics in agent-based models}
\author[a,1]{Nata\v sa Djurdjevac Conrad}
\author[b,a]{Nhu Quang Vu}%
\author[a]{Sören Nagel}

\affil[a]{Zuse Institute Berlin, Germany}
\affil[b]{Department of Mathematics and Computer
Science, Institute of Computer Science,Freie Universität Berlin}
\affil[1]{Corresponding author: natasa.conrad@zib.de}

\begin{document}

\maketitle
\begin{abstract}
The rise of digital social media has strengthened the coevolution of public opinions and social interactions, that shape social structures and collective outcomes in increasingly complex ways. Existing literature often explores this interplay as a one-directional influence, focusing on how opinions determine social ties within adaptive networks. However, this perspective overlooks the intrinsic dynamics driving social interactions, which can significantly influence how opinions form and evolve. In this work, we address this gap, by introducing the co-evolving opinion and social dynamics using stochastic agent-based models. Agents’ mobility in a social space is governed by both their social and opinion similarity with others. Similarly, the dynamics of opinion formation is driven by the opinions of agents in their social vicinity. We analyze the underlying social and opinion interaction networks and explore the mechanisms influencing the appearance of emerging phenomena, like echo chambers and opinion consensus. To illustrate the model's potential for real-world analysis, we apply it to General Social Survey data on political identity and public opinion regarding governmental issues. Our findings highlight the model's strength in capturing the coevolution of social connections and individual opinions over time.
\end{abstract}

\maketitle


\section{\label{sec:Intro}Introduction}

Availability of large amounts of data from online social media has intensified research on several longstanding questions: How do individuals form opinions within their social environment? Which factors lead to opinion polarization? How does the spread of misinformation affect opinion formation and collective decision-making? Extensive studies on these topics have been developed in the last decades, resulting in a variety of approaches for understanding social mechanisms and opinion dynamics \cite{lorenz2007,peralta2022OpinionReview,Loreto2017OpinionReview,social_physics_review}. These are ranging from model-driven approaches, that create formal mathematical models, to data-driven approaches, that analyze empirical data. However, there is still a large gap between these two directions, since many formal models fail to capture real-world mechanisms and rarely connect with empirical data. Closing this gap requires novel formal models that can better capture the rich behavior of real-world social systems.

Agent-based models (ABMs) have been shown to be a powerful tool for studying opinion and social dynamics \cite{deffuant2000,schweitzer2003book,lorenz2007,Castellano2009,helfmann2023}. Starting with microscopic action and interaction rules of individual agents, these models can capture emergence of large-scale phenomena, such as opinion consensus, in which all agents share the same opinion. Alternatively, such systems can reach a state of fragmentation, i.e. appearance of echo chambers, where like-minded agents are grouped together. Pioneering ABMs for opinion dynamics \cite{hegselmann2002,deffuant2000} assume full-connectivity, where every agent can interact with all other agents. However, in the context of real-world social systems this is not a realistic assumption, as interactions between individuals are often guided by their social networks. Introducing complex networks to opinion dynamics ABMs yields deeper insights into its dynamics connecting topological network properties to possible stable states of the system \cite{gabbay2007,SvenEckehard2019,KanPorter2023}. Particularly relevant in this context are the so-called adaptive (or co-evolving) network models \cite{HolmeNewman2006,kozma2008,ChaosAdaptiveNet17,PhilippPRX21,KurthsEtAl2023Review}, where network structure evolves through a rewiring mechanism based on agents’ opinions, such that agents form connections to agents holding similar opinions and disconnect from those they disagree with. This mechanism is shown to hinder the system's ability to reach global consensus, but rather leads to fragmentation \cite{ChaosAdaptiveNet17}. A key question studied with such models is: Do social network connections shape individual opinions, or conversely, are people with similar opinions more likely to connect? This question is particularly relevant in online social media, where new information continuously reshape both the structure of social connections, as well as the opinion distribution \cite{Holme14SocMedia}. Dynamic interplay between opinion and social dynamics is a crucial aspect of understanding opinion formation and social grouping in today's online world. 

Although the coevolution of opinion and social network dynamics has been studied in various contexts, such as in coevolutionary games addressing social dilemmas \cite{evolutionary_games_review, Coevolutionary_games_mini_review}, most such ABMs neglect incorporating the intrinsic social dynamics of agents, a key factor influencing social interaction patterns. In this work, we aim to address this gap by introducing a social space where "mobile agents" \cite{starnini2016emergence,DjCKoDj22} move governed by stochastic dynamics and form a time-evolving social interaction network \cite{HolmeSaramaki2012, Holme2014}.  Agents’ mobility is governed by both their social and opinion similarity with others; and similarly, opinion dynamics is driven by the opinions of agents in their social vicinity. The co-evolutionary dynamics of the system are driven by this feedback loop between movement in the social space and opinion changes. We formulate our model with a system of coupled stochastic differential equations (SDEs) and study which factors drive the emergence of consensus, or alternatively echo chambers, in the process of opinion formation and how these organize within the social system. In particular, we analyze the underlying social and opinion interaction networks, with respect to the feedback loop that guides their evolution in time. Motivated by real-world scenarios, such as political elections and policy changes, our focus is on the transient model dynamics, where metastable network clusters exhibit rich dynamics. Estimating the number of such clusters and their dynamics has been a topic of several recent studies \cite{kozma2008,garnier2017,ChaosAdaptiveNet17}. We compare the model’s behavior with empirical data from General Social Survey \cite{davern2024gss}, and use political views and affiliations to define a variant of the political spectrum. In this context, we explore opinion distributions on two government related issues and observe that for these questions social influence dominates the opinion dynamics. If such system has extremely large number of agents, performing ABM simulations and model calibration becomes infeasible. We show that in these cases, mean-field models offer a good approximation of the system with reduced computational costs.

This article is organized as follows. In Section \ref{sec:Model} we present the agent-based model for the co-evolving opinion and social dynamics, together with its main dynamical characteristics. Next, in Section \ref{sec:CoevolvingNetworks}, we introduce the two underlying interaction networks and study how their coevolution can lead to emerging phenomena. In Section \ref{sec:MeanField} we present the mean-field model and compare it to the original ABM. We show how our model can be applied to empirical data-sets in Section \ref{sec:GSSResults}. Finally, in Section \ref{sec:Conclusions} we derive our conclusions and possible future directions.

\section{\label{sec:Model}An agent-based model for opinion and social dynamics}

For a fixed number of $N$ interacting agents, we consider a model of co-evolving opinion and social dynamics, as previously introduced in Djurdjevac Conrad et.al. \cite{DjCKoDj22}. At time $t\in[0,T]$ each agent $k = 1,\ldots,N$ holds an opinion $\theta_k(t)\in \R^m$ and has a position in a social space $x_k(t) \in \R^d$. Opinions and positions of agents change in time, but for brevity we omit the dependence on time in our notation and write $x_k$ and $\theta_k$ instead of $x_k(t)$ and $\theta_k(t)$. 
Defined in this way, $\theta_k = (\theta_k^1, \ldots, \theta_k^m)$ denotes opinions toward $m$ different topics, and the sign structure of each element $ \theta_k^l\in\R$ can be interpreted as the agents standing towards this particular topic $l$ (its stance), e.g. support or oppose \cite{PhilippPRX21}. The position $x_k$ represents a point in an abstract social space, such that the distance between two agents indicates their social similarity. Central to our model is a feedback loop between opinion and social dynamics, such that the opinions of agents are not only influenced by other opinions but also by the social proximity of other agents and vice versa, the position of agents in a social space are governed by both the position of other agents and their opinions. Although the original model \cite{DjCKoDj22} was formulated such that it could incorporate multiplicative noise and higher order interactions, in the following we  consider the systems with additive noise and pairwise agent interactions. The dynamics of the agents is described by the following system of stochastic differential equations (SDEs) \cite{DjCKoDj22}
\begin{equation}\label{eq:ABM}
\begin{split}
    dx_k &= \frac{1}{N}\sum^N_{j=1}U(x_k,x_j,\theta_k,\theta_j) dt + \sigma_{sp} dW_k^{sp}(t), \\
    d\theta_k &=\frac{1}{N}\sum^N_{j=1}V(x_k,x_j,\theta_k,\theta_j) dt + \sigma_{op} dW_k^{op}(t), 
\end{split}
\end{equation}
where $U$ defines the \textit{social (spatial) interaction map} and $V$ is the \textit{opinion interaction map}, $\sigma_{sp}, \sigma_{op} >0$ are diffusion coefficients and $W_k^{sp}(t) \text{ and } W_k^{op}(t)$ are independent Brownian motions. Like in other non-deterministic opinion dynamics models  \cite{pineda2009,schweitzer2000,pineda2013,su2017,GoddardPavliotis2021}, the noise is introduced here to account for external influences, uncertainties in the system. The choice of the spatial and opinion interaction functions determines the impact of agents' opinions and positions on other agents' opinions and positions. 

Various possible choices of interaction maps have been studied in the literature \cite{proskurnikov2018tutorial,Gomes2023,helfmann2023}. Here, we consider a bounded confidence model \cite{hegselmann2002,deffuant2000} that only allows interactions with other agents that are within a specific distance (called interaction radius or confidence bound) from each other. The rationale behind this modeling choice is that agents who are very distant in social space, meaning they have low social similarity, might possess conflicting attitudes and social norms and therefore may lack motivation to engage with each other. 

Since in our model there are two co-evolving processes, we introduce two confidence bounds, i.e. the \textit{spatial interaction radius} $R_{sp}$ and the \textit{opinion interaction radius} $R_{op}$. We assume that the movement of an agent in a social space is influenced by social interactions with other agents whose position is within $R_{sp}$ of that of that agent. Similarly, the opinion of an agent is influenced by social interactions with other agents that are at most $R_{op}$ distance (in social space) from that agent. We introduce two different interaction radii to reflect the fact that not the same level of social similarity is needed for agents to influence their opinions as it is needed to influence their social positions. For example, we can consider scenarios where $R_{sp} > R_{op}$ in which only the agents of the "inner social circle", i.e. the closest peers, can influence the opinion formation process. Conversely, scenarios where $R_{sp} < R_{op}$ reflect situations where interactions outside of the usual social environment are impacting the process of opinion formation, e.g. due to online communications that can be made across spatial and sociodemographic constrains. The focus of existing literature has largely been on exploring 
the cases where $R_{sp} = R_{op}$ \cite{Gomes2023,PhilippPRX21} and extensions to cases when $R_{sp} \neq R_{op}$ have have not been studied.

In the following, we consider a particular choice of the pair-interaction maps $U,V$ defined by 
\begin{equation}\label{eq:UV}
\begin{split}
    U(x_k,x_j,\theta_k,\theta_j)&:= \beta\cdot1_{[0,R_{sp}]}(\Vert x_k-x_j\Vert)\cdot \text{sgn}(\theta_k\cdot\theta_j)\cdot(x_j-x_k), \\
    V(x_k,x_j,\theta_k,\theta_j)&:= \alpha\cdot1_{[0,R_{op}]}(\Vert x_k-x_j\Vert)\cdot(\theta_j-\theta_k), 
\end{split}
\end{equation}
for the \textit{opinion influence strength} $\alpha$ and the \textit{social influence strength} $\beta$. Parameters $\alpha$ and $\beta$ govern the adaptability of the system by regulating the impact of opinion and social dynamics on the agents' opinion and position in a social space, respectively. More precisely, higher values of $\alpha$ and $\beta$ indicate that agents compromise stronger towards their neighbours when updating their opinions and positions. This behavior affects the speed of convergence of the system, which is why $\alpha$ and $\beta$ are commonly referred to as ‘convergence parameters’ \cite{kozma2008,KanPorter2023}. The feedback loop in the social dynamics is introduced in \eqref{eq:UV} in the following way: 
for two agents $j, k$ that have the same stance towards a topic, i.e. $\text{sgn}(\theta_j)=\text{sgn}(\theta_k)$, there exists an attractive force between these agents in the social space and agents will mutually reinforce each other driven by their social proximity. Similarly, there is a repulsive force  in the social space between agents with opinions of a different sign. In a real-world context this may correspond to social distancing between people of different opinions and when the distance is small (i.e. less than $R_{sp}$) it can result in effects like unfriending \cite{KanPorter2023}. The feedback loop in the opinion dynamics  \eqref{eq:UV} is defined such that opinions evolve under the influence of the social interactions (which are driven by agents' social proximity) and homophily, i.e. two agents can interact if they are close enough in a social space and if they interact, their opinions become more similar. This concept is derived from the phenomenon of group polarization \cite{myers1976,isenberg1986}, that is characterized by the reinforcement of individual opinions within a group of individuals sharing similar views, resulting in a shift towards more extreme positions or viewpoints. Similar mechanisms have been explored in other models \cite{HolmeNewman2006,KimuraHayakawa2008,PhilippPRL2020,PhilippPRX21}.

\subsection{Impact of different time-scales}

Microscopic interactions between agents that are driven by co-evolving opinion and social dynamics in our model, can lead to different macroscopic patterns. With respect to the opinion states, agents may reach an agreement and form an opinion consensus, sharing very similar opinions. Alternatively, fragmented clusters of opinions can appear, where in a case of two opinion clusters, the emerging phenomena is called opinion polarization. Similarly, clustering of agents can be observed in the social space, where we distinguish between social "consensus", when all agents are grouped together in the social space, and social segregation, with the existence of a few, up to many groups that are separated in social space. Through the interplay between these two co-evolving dynamics in our model, we can additionally observe a very rich behavior. These can emerge as a consequence of a competition between the cluster formation in the social space and homophilic opinion dynamics, that drive the system towards a consensus state. The final state of the system is determined by the ratio between the time-scale of the social and opinion dynamics. If the social dynamics is slow compared to the opinion dynamics, a global opinion consensus emerges as agents with similar opinions gradually align their views through interactions. Conversely, faster social dynamics leads to a fragmented social landscape, where each of the resulting clusters reaches an opinion consensus locally. Fragmentation happens fast, through the social distancing of agents with conflicting opinions that aim to reduce their social influence on each other. We study how the two co-evolving processes influence the nontrivial relations between the opinion formation and social structures.

\subsection{\label{subsec:EmergingPatterns}Emerging opinion and social patterns}
We explore the effect of influence strength parameters $\alpha$, $\beta$ on the appearance of emerging patterns in both social and opinion space. For technical simplicity and convenient illustrations, we study the behavior of the system with $N=100$ agents and the case where the social space is two-dimensional ($d=2$) with values from $[-0.25, 0.25]^2$ and the opinions are expressed on one topic ($m =1$) with values from the range $[-1,1]$. Additionally, we choose $R_{op} = R_{sp} = 0.15$ and $\sigma_{op} = \sigma_{sp}= 0.05$. Initially, all agents are placed uniformly at random inside the social space and their initial opinions are distributed uniformly in the interval $[-1,1]$. We focus on transient regime and run simulations until $T = 2.5$. Movements and opinions of agents are obtained using a standard Euler—Maruyama scheme \cite{kloeden1992higher}, with a time-step size $\Delta t= 0.01$.

\begin{figure}
    \centering
    \includegraphics[width=.5\textwidth]{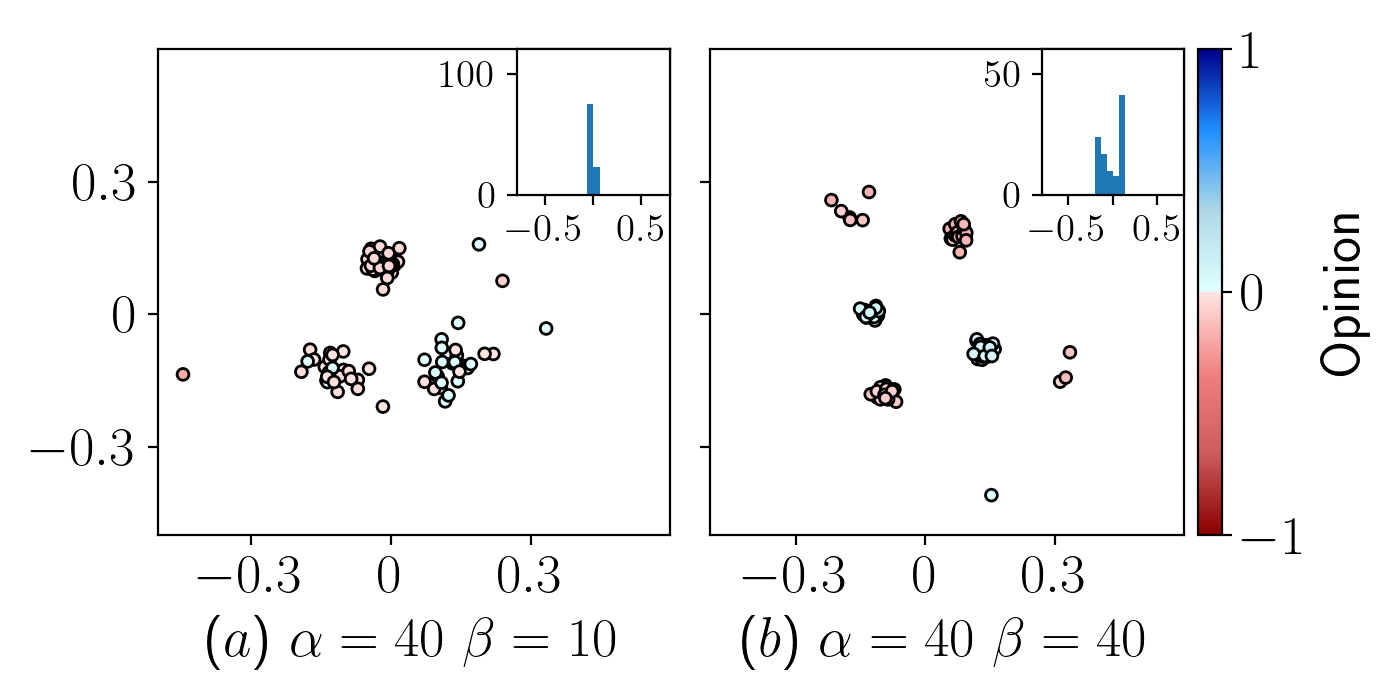}
    \caption{The agents position in the social space for different values of $\beta$ at time $T=2.5$. The colors represent the opinions of the agents. The insets show the opinion distribution. Note the different scales of the histograms. The remaining parameters: $N=100$, $R_{op}=R_{sp}=0.15$, $\sigma_{op} = \sigma_{sp}=0.05$.}
    \label{fig:ChangeBeta}
\end{figure}

The choice of parameters $\beta$ and $\alpha$ affects the magnitude of the forces acting in the social and opinion dynamics, respectively. In Fig.~\ref{fig:ChangeBeta} we show snapshots from one realisation at time $T=2.5$ for a fixed $\alpha = 40$ and different values of $\beta=10$ and $\beta =40$. The positions of agents correspond to their position in the social space and colors to their opinion. The color map  contains a clear separation between positive (blue) and negative (red) values of opinions to visualise the discontinuity at $\theta=0$. For $\beta=10$ and the ratio $\frac{\alpha}{\beta} = \frac{4}{1}$, the impact of the social dynamics is not very strong and the homophilic opinion interactions dominate. The resulting opinion distribution shows that opinions are concentrated around zero. In a social space, agents are arranged in a few loose clusters with heterogeneous opinions, shown in Fig.~\ref{fig:ChangeBeta}(a). These clusters are prone to changes due to the alternating forces of repulsion and attraction between agents. We study evolution of this system on a longer time-scale in Section \ref{subsec:transient_dynamics}. As the social influence increases to $\beta = 40$, social interactions gain stronger impact. Since $\alpha = 40$, both dynamics are happening on the same time-scale. Thus, the resulting opinion distribution, although being bimodal, has values that are relatively close to $0$. Bi-modality arises from the cluster formation in the social space, where agents holding similar opinions of the same sign are grouped together, see Fig.~\ref{fig:ChangeBeta}(b). Once the clusters are formed, interactions between agents of different stances are possible, due to $R_{op} = R_{sp} = 0.15$, but are short-lived as they cause repulsion and further separation in the social space. Eventually clusters become well separated and stay metastable, i.e. they are stable for a long time with rare transitions under the impact of noise.

\begin{figure}
    \centering
    \includegraphics[width=.5\textwidth]{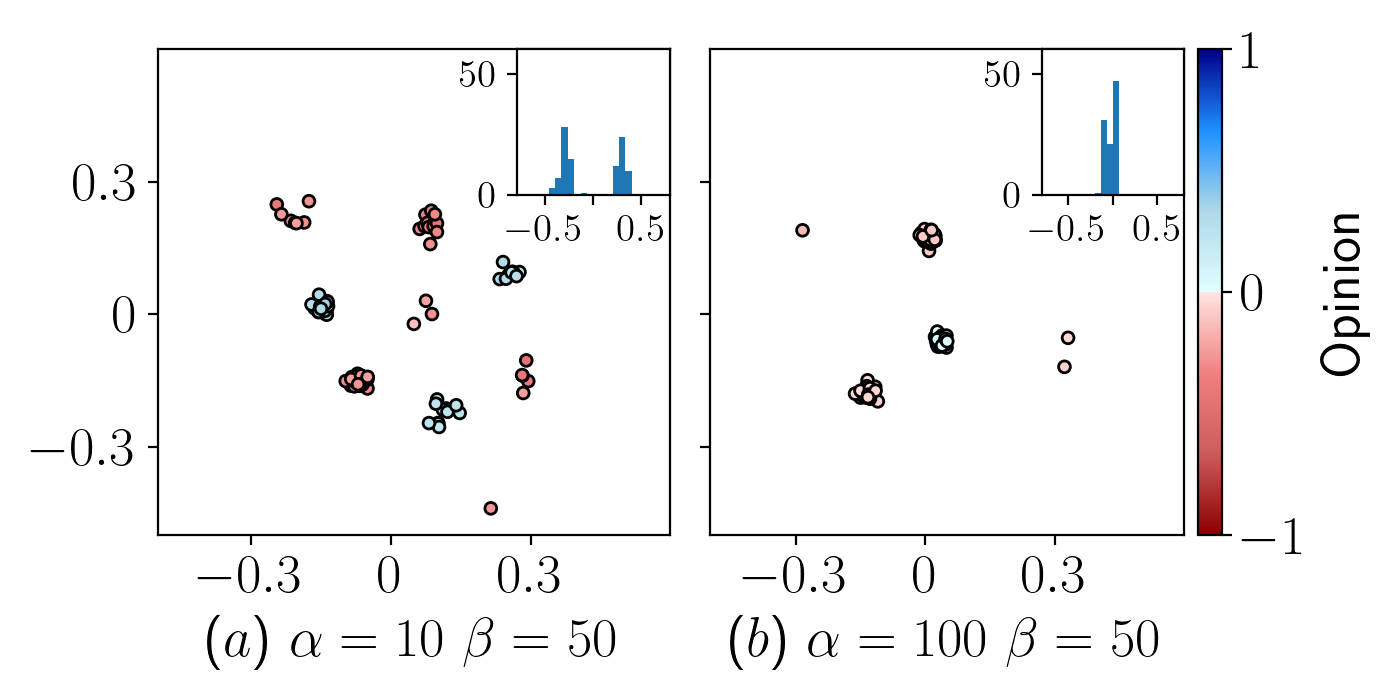}
    \caption{The agents position in the social space for varying overall coupling strength at time $T=2.5$. The colors represent the opinions of the agents. The insets show the opinion distribution.  The remaining parameters: $N=100$, $R_{op}=R_{sp}=0.15$, $\sigma_{op} = \sigma_{sp}=0.05$.}
    \label{fig:ChangeAlpha}
\end{figure}

When the opinion influence strength is weak $\alpha=10$ compared to social influence $\frac{\alpha}{\beta} = \frac{1}{5}$, social interactions dominate the dynamics, see Fig.~\ref{fig:ChangeAlpha}(a). This leads to a rapid separation of agents into many clusters with opposing opinions. The resulting opinion distribution shows a strong polarization, with distinct groups of strictly positive and negative opinions. Increasing $\alpha$ results in a stronger attraction between agent's opinions, amplifying the relative effect of homophily. For a high $\alpha = 100$ and $\frac{\alpha}{\beta} = \frac{2}{1}$, only very few dense clusters emerge in the social space, each having homogeneous opinions with the same stance. Similar to the case $\alpha = 40, \beta = 10$, the opinion distribution of the whole system is centered around $\theta=0$, see Fig.~\ref{fig:ChangeAlpha}(b). This is because the fast opinion convergence towards the mean opinion (driven by a high value of  $\alpha$), counteracts the repulsive forces in the social space before these can separate agents into clusters with opposing opinions.

A crucial question is on how many clusters emerge in the system when it doesn't reach a global consensus. In the case of deterministic and stochastic Hegselman-Krause model, analytical results using linear stability analysis of the mean-field model, can be used to estimate the number of clusters and the time to cluster formation \cite{garnier2017}. Studies that include some types of social networks suggest that the number of clusters $M$ roughly follow the law $M\sim \frac{1}{2R}$, where  $R$ is an interaction radius \cite{kozma2008,hegselmann2002}. Cluster dynamics becomes very complex in more general models, as additional factors like mutation rates can significantly influence cluster formation \cite{ChaosAdaptiveNet17}. Our model, due to the feedback loop, exhibits even rich cluster dynamics that strongly depends on various model parameters. Analytical results on estimating the number of clusters in our model will be the topic of future research.

\section{Dynamics of co-evolving networks}\label{sec:CoevolvingNetworks}
Through the feedback loop, interactions between agents drive both the social and opinion dynamics. These interactions are not static, they are changing in time and form time-evolving (also called temporal or dynamic) interaction networks \cite{HolmeSaramaki2012, Holme2014}. Corresponding social and opinion interaction networks, that we denote as $G^{sp}(t)$ and $G^{op}(t)$ respectively, co-evolve in time \cite{gross2008} and they can be represented by time-dependent adjacency matrices. In particular, $A^{sp}(t)$ captures the social interactions 
\begin{align}\label{def:A_sp}
A_{jk}^{sp}(t):=\begin{cases}
1, &\text{if } \Vert x_k(t)-x_j(t)\Vert \leq R_{sp} \\
0, &\text{else },
\end{cases}
\end{align}
and $A^{op}(t)$ captures the opinion interactions 
\begin{align}\label{def:A_op}
A_{jk}^{op}(t):=\begin{cases}
1, &\text{if } \Vert x_k(t)-x_j(t)\Vert \leq R_{op} \\
0, &\text{else },
\end{cases}
\end{align}
at specific time $t$. As discussed in Section \ref{sec:Model}, both networks $G^{sp}(t)$ and $G^{op}(t)$ reflect social similarity as they depend on the proximity of agents in the social space.  
Model \eqref{eq:ABM} can now be formulated with respect to the co-evolving networks as
\begin{equation}\label{eq:ABM_Net}
\begin{split}
    dx_k &= \frac{\beta}{N}\sum^N_{j=1} A_{jk}^{sp}(t)\cdot sgn(\theta_k\cdot\theta_j)\cdot(x_j-x_k) dt + \sigma_{sp} dW_k^{sp}(t), \\
    d\theta_k &=\frac{\alpha}{N}\sum^N_{j=1}A_{jk}^{op}(t)\cdot(\theta_j-\theta_k) dt + \sigma_{op} dW_k^{op}(t).
\end{split}
\end{equation}
Structures of the two networks play a crucial role in shaping opinion and social dynamics. Of particular interest are topological network clusters, i.e. densely connected subgraphs that are loosely connected to the rest of the network. Additionally, we refer to distinct connected components also as topological clusters. Tracking the evolution of network clusters offers a way to study how the system's structure changes over time. Agents tend to form clusters based on their social proximity and opinion similarity. Dense connections within topological clusters of opinion interaction network $G^{op}$ impose frequent opinion interactions between the agents, leading to local consensus. Resulting clusters are metastable, i.e. they tend to stay stable over long periods of time  with rare transitions under the influence of noise. However, before reaching a (meta-)stable state, the network undergoes interesting transient dynamics that we explore next. 

\subsection{Transient dynamic\label{subsec:transient_dynamics}}
 
\begin{figure}
    \centering
    \includegraphics[width=.6\textwidth]{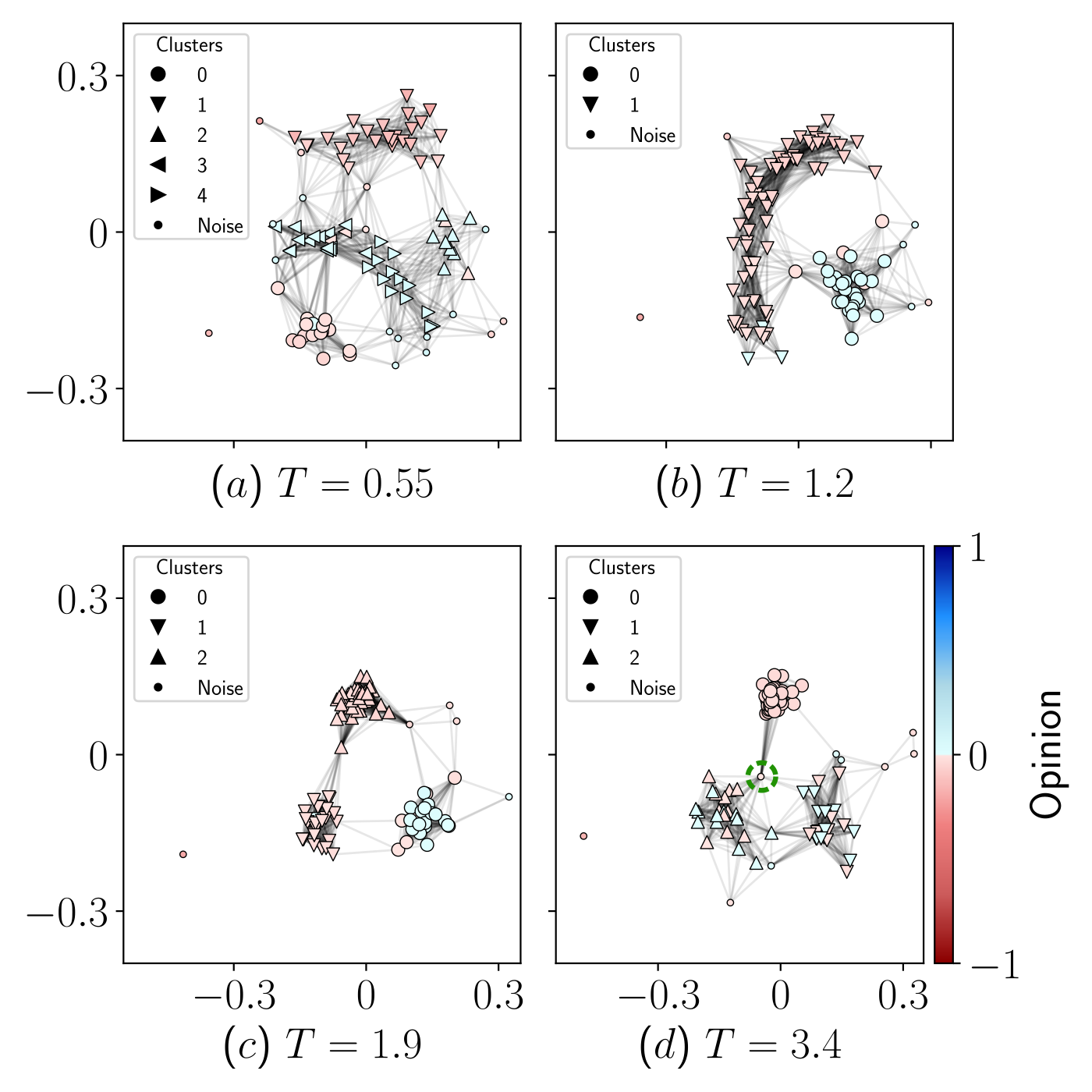}
    \caption{Snapshots of a single realization highlighting interesting cluster evolution. Shape of a node encodes a cluster it belongs to and color indicates the opinion state. The lines indicate edges of the interaction network $G^{op} = G^{sp}$, for $R_{op}=R_{sp}=0.15$. The remaining parameters: $N=100$, $\alpha=40$, $\beta=10$, $\sigma_{op} = \sigma_{sp}=0.05$.}
    \label{fig:Network_Dynamics}
\end{figure}

Long-term evolution of our system leads to a formation of either one consensus state or several isolated clusters with local consensus. Here, we explore an example of the transient dynamics preceding the stable state, for the same choice of parameters as in Fig.~\ref{fig:ChangeBeta}(a). In Fig.~\ref{fig:Network_Dynamics}, we plot four snapshots from one realization, highlighting the plethora of dynamical situations that can occur in the transient. We cluster the network at each time using a density-based clustering algorithm (HDBSCAN) \cite{mcinnesHdbscanHierarchicalDensity2017}, that takes only the position in the social space, but not the opinions of the agents into account. Shape of each node indicates the cluster it belongs to and its color corresponds to the opinion state at that time. Note, that the labels for the clusters are chosen independently for each time step. 

 Already at $T=0.55$ (Fig.~\ref{fig:Network_Dynamics}(a)) agents get organized into four main clusters, with several nodes that do not deterministically belong to any of these (labeled as "Noise"). Clusters 0 and 1, consisting of agents that are predominately red, are "separated" by the clusters 2, 3 and 4, whose agents are predominately blue. Since these groups have opinions of different signs, there is a repulsive force in the social space between them. Nodes in cluster 3 are connected to many of the nodes in both cluster 0 and 1, creating a situation where there is a significant number of edges between clusters of opposing opinions for a considerable amount of time. These prolonged interactions increase the probability that agents in cluster 3 switch their opinion sign and they become absorbed by much bigger clusters 1 and 3, forming a single large cluster at time $T = 1.2$, see Fig.~\ref{fig:Network_Dynamics}(b). During the same time interval, driven by attraction force due to the similar opinions, cluster 4 and cluster 2 merge into one cluster. However, the structure of cluster 1 changes under the influence of mixed opinions of its nodes, such that these separate into 2 clusters at $T = 1.9$. Then, at time $T = 3.4$, under the persistence of intra-cluster interactions and all opinions being relatively close to $0$, the agents in cluster 1 switch from predominantly red to mixed opinions, Fig.~\ref{fig:Network_Dynamics}(d). This consensus on the discontinuity and the presence of noise, result in both attractive and repulsive interactions being present. In such regime, the network clusters are not stable and various (sometimes random) effects might determine the network evolution. We observe this on the example of a highlighted agent in Fig.~\ref{fig:Network_Dynamics}(d). This node was originally assigned to cluster 1, but as it connects to many agents in cluster 2 (which are all of the same opinion), it eventually leaves cluster 1 and becomes assigned to cluster 2. Studying the influences that can lead to such scenarios can help understanding how a radicalization of individual agents can take place.

\subsection{Impact of different interaction radii}\label{subsec:RadiusVariation}
In standard bounded confidence models, the interaction radius  plays a crucial role in shaping the dynamics of opinion formation. When the radius is large, the network is densely connected and agents can interact with many other agents, resulting in a global consensus. Conversely, for a small interaction radius the network separates into clusters with distinct opinion groups. We introduced two distinct interaction radii that determine the structure of networks $G_{sp}$ and $G_{op}$ and the co-evolution of opinion and social dynamics. In previous sections, we looked at simulations where $R_{op} = R_{sp}$, which is the case typically studied in the literature. Here, we explore the impact of differing values of social and opinion interaction radii on the formation of opinion and social patterns. 

\begin{figure}
    \centering
    \includegraphics[width=.5\textwidth]{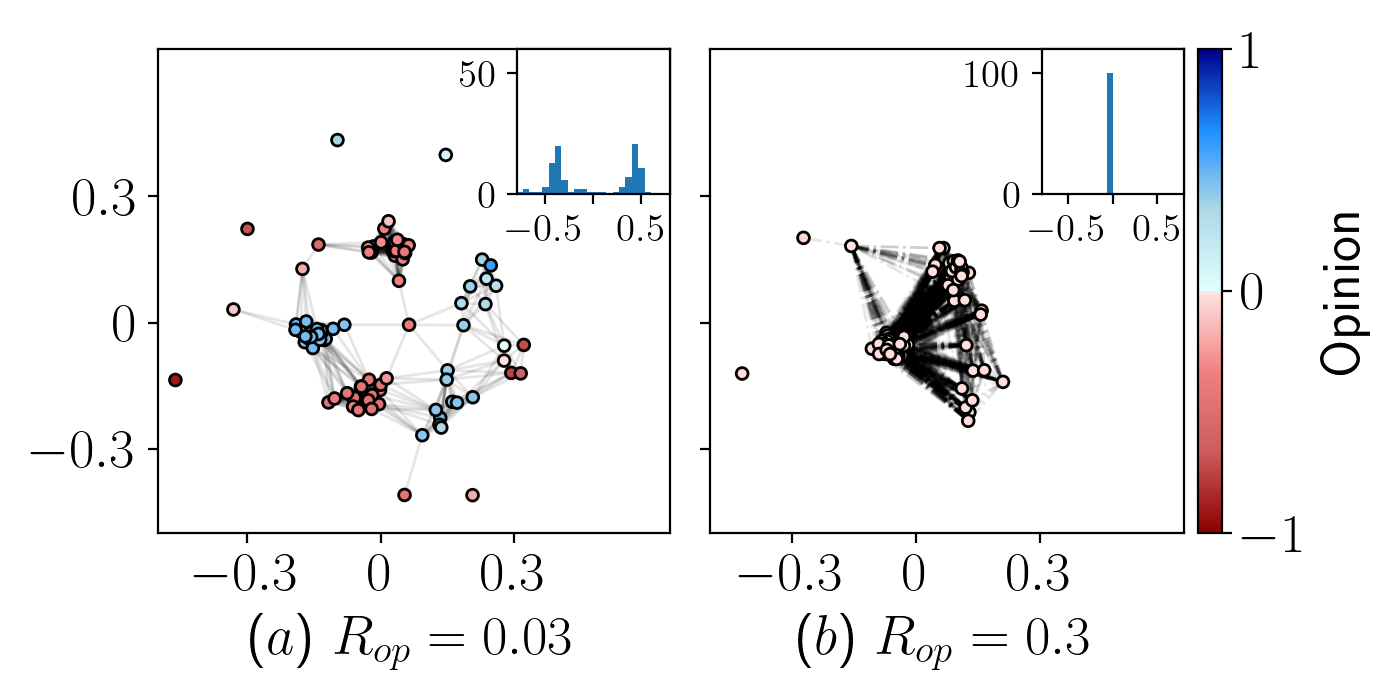}
    \caption{The agents position in the social space for different values of $R_{op}$ at time $T=2.5$. (a) For the choice of $R_{op} = 0.03$ we plot network $G_{sp}$; and (b) for $R_{op} = 0.3$ we plot network $G_{op}$, with dashed lines. The colors represent the opinions of the agents. The insets show the opinion distribution. Note the different scales of the histograms. The remaining parameters: $N=100$, $R_{sp}=0.15$, $\sigma_{op} = \sigma_{sp}=0.05$, $\alpha = 40$ and $\beta= 10$.}
    \label{fig:opinion_radius}
\end{figure}

We begin by studying the case when the spatial interaction radius $R_{sp}$ is larger than the opinion interaction radius $R_{op}$. In Fig.~\ref{fig:opinion_radius}(a) we show a snapshot at time $T = 2.5$  from a simulation with parameters $R_{sp} = 0.15$ and $R_{op} = 0.03$. The corresponding interaction network $G_{sp}$ is divided into several interconnected clusters with locally similar opinions. However, despite social interaction within and across spatial clusters, there is almost no opinion interaction between clusters, due to a very small $R_{op}$. Topological clusters of the opinion interaction network $G_{op}$ coincide only with the cores of clusters from $G_{sp}$, but because of the small radii $R_{op}$ there are almost no edges between distinct clusters. This leads to a polarized opinion distribution, with each cluster exhibiting some internal diversity in opinions. 

Next, when $R_{sp} = 0.15 < R_{op} = 0.3$, due to the very large opinion interaction radius a global opinion consensus is reached, see Fig.~\ref{fig:opinion_radius}(b). However, at $T = 2.5$ there are still disconnected clusters in network $G_{sp}$ that will stay separated until the threshold $R_{sp}$ of social similarity is crossed, after which the attraction forces will guide the nodes of $G_{sp}$ to one connected component. Compared to scenarios where $R_{op} = R_{sp} = 0.15$, we see that for fixed $R_{sp}$, the choice of opinion interaction radius has the same impact as in standard bounded confidence models. In particular, for cases like the one shown in Figure \ref{fig:ChangeBeta}(a), where the opinions are grouped around zero, there is a mixture of attractive and repulsive forces, that reduces density of cluster(s) and thus, the speed of convergence. We show similar experiments for the changing values of $R_{sp}$ in the Appendix in Fig.~\ref{fig:ChangingR_sp}.

\subsection{Measuring network assortativity}
Here we explore how the combined analysis of the opinion distribution and network structures can be used as an indicator of polarization in our model. On the one hand, there is a rich literature on quantifying the diversity of opinion distributions and on the other hand, there are various methods for analyzing structural network properties. Our focus, however, is on quantifying the interplay between both aspects and examining the temporal evolution of such a measure.\\
We achieve this by employing the global network assortativity coefficient  \cite{NewmanAssortativity02, NewmanAssortativity03}, which measures the tendency of nodes to be connected to similar nodes. For analysis of real-world networks, node similarity is often defined with respect to node degree, such that the assortativity is calculated as the Pearson correlation coefficient between the degrees of connected nodes. In the context of our analysis, we consider node similarity based on both the opinion values and structure of the network $G_{sp}$. In particular, we define the global assortativity as
\begin{equation}\label{eq:assortativity}
    r = \frac{\sum_{j, k = 1}^{N}{A_{jk}^{sp}(\theta_j -\bar{\theta})(\theta_k-\bar{\theta})}}{\sum_{j=1}^N{d_j^{sp}(\theta_j-\bar{\theta})^2}},
\end{equation}
where $\bar{\theta} = \frac{1}{2m}\sum_{j=1}^N d_j^{sp} \theta_j$ is the mean opinion value of $\theta$ weighted by node degree $d^{sp}$ of the interaction network $G^{sp}$ with $m$ edges. 
Positive global assortativity values $r>0$ indicate that nodes with similar opinions are more likely to be connected in the network $G_{sp}$, while negative values suggest that connections between nodes having distinct opinions are more likely. Additionally, the global assortativity coefficient is normalized such that for $r = 1$ network is perfectly assortative, for $r = 0$ network is non-assortative (or well mixed) and for $r = -1$ network is completely disassortative. Defined in this way, global assortativity accounts for both social dynamics through the evolution of network $G^{sp}$ and opinion dynamics via changing opinion distributions, thus incorporating the entire feedback loop. 

\begin{figure}
    \centering
    \includegraphics[width=0.5\textwidth]{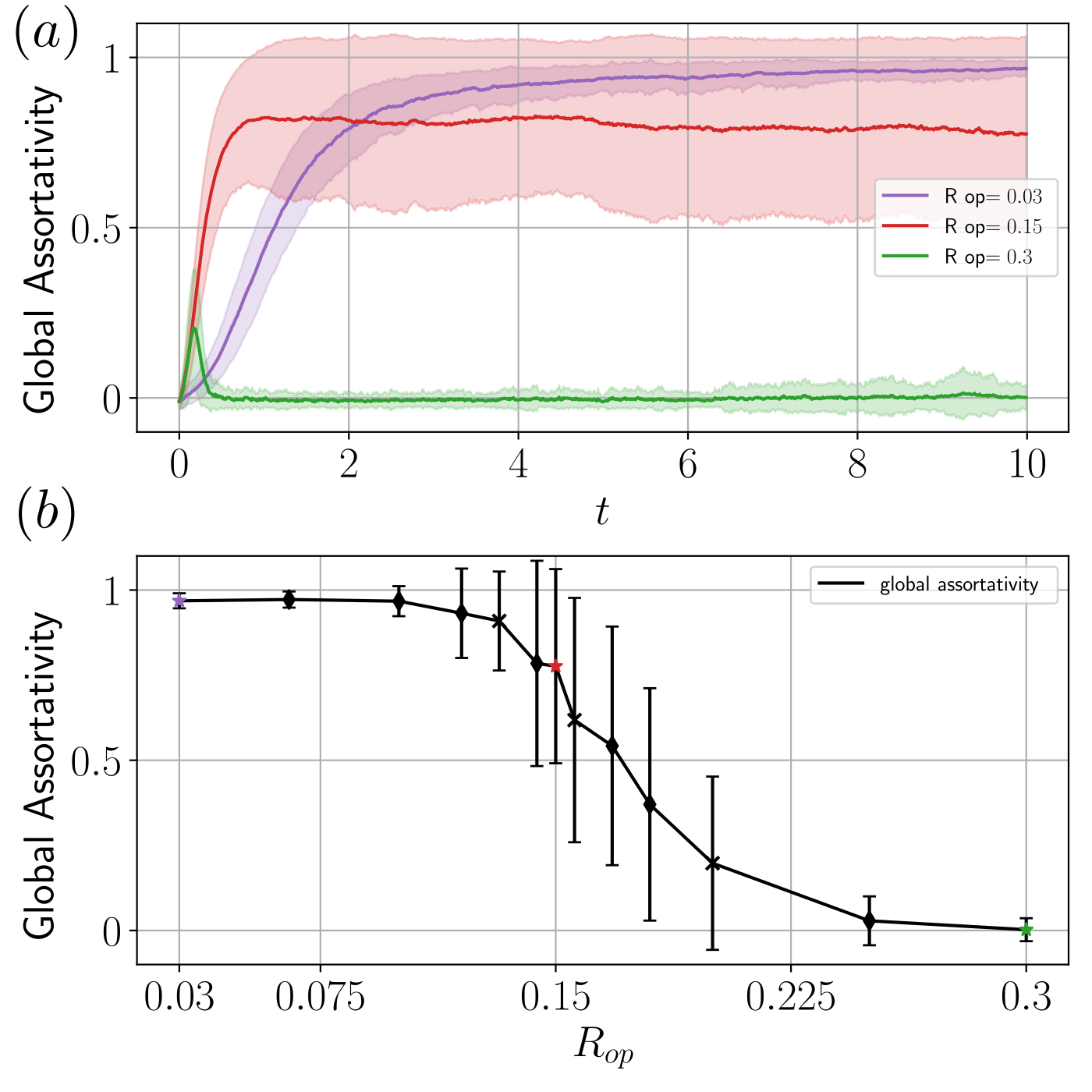}
    \caption{(a) Comparison of the temporal evolution of the mean global assortativity value averaged over 100 simulations and standard deviation (shaded area). (b) The mean value of global assortativity at $T=10$ for several values of $R_{op}$, averaged over 100 realizations and standard deviation (shown by the error bars). The $R_{op}$ values shown in (a) are marked in (b) by stars and in their respective color. In the intermediate regime the assortativity shows bimodality, these $R_{op}$ values are marked by 'x'. Other Parameters as in Fig.~\ref{fig:opinion_radius}.}
    \label{fig:global-assortativity}
\end{figure}

In order to study the impact of the opinion interaction network, we compare global assortativity values for different values of opinion interaction radius $R_{op}$. In  Fig.~\ref{fig:global-assortativity}(a) we plot the temporal evolution of the mean global assortativity value over $100$ simulations and the standard deviation for: 
\begin{enumerate}
    \item[(i)] a small value of $R_{op} = 0.03$, such that $R_{op}<R_{sp}$; 
    \item[(ii)] a large value of $R_{op} = 0.3$, where $R_{op}>R_{sp}$;
    \item[(iii)] equal interaction radii $R_{op} = R_{sp} = 0.15$.
\end{enumerate}
We observe two main regimes based on the values of $r$. First, for the parameters from (i) where $R_{op}<< R_{sp}$, the network becomes perfectly assortative with values converging to $1$. Indeed, as shown in Fig.~\ref{fig:opinion_radius}(a), already at $T = 2.5$, nodes form clusters in $G_{sp}$ with locally homogeneous opinions. Second, for parameters from (ii), i.e. for $R_{op}>R_{sp}$, the network is in a well mixed regime with $r \approx 0$ after the initial quick reorganization of the system. As we can see in Fig.~\ref{fig:opinion_radius}(b), due to global consensus, there's no significant correlation between the opinions of connected nodes. In the intermediate regime, where $R_{op} = R_{sp} = 0.15$, the mean assortativity shows bimodality.\\
To gain more insights into the transition between the high and low assortativity regime, we investigate the behaviour for several values of $R_{op}$ in Fig.~\ref{fig:global-assortativity}(b). For each value of $R_{op}$ we again average over $100$ realizations and illustrate the standard deviation using error bars. The choice of small values of $R_{op}$, results in high mean assortativity values. As we approach the case of $R_{op} = R_{sp} = 0.15$, the mean assortativity decays, since the increase in the opinion radius reduces the separation between clusters in space and opinions and creates edges (interactions) across different clusters. Notably, in this intermediate regime, the assortativity exhibits a high degree of uncertainty, indicating the coexistence of both highly assortative and well-mixed states. We explore such scenarios in the Appendix \ref{app:futher_results} in Fig.~\ref{fig:bimodality}. As described in Fig.~\ref{fig:global-assortativity}(a), for large values of $R_{op}$, agents can influence each other even when they are not connected in $G^{sp}$, so the population comes closer to forming a consensus, which results in reduced assortativity that for higher values reaches the well-mixed state.

\subsection{\label{subsec:SocialSpace} Introducing a shape of the social space}

Scenarios considered so far demonstrate a rich dynamical behavior of the system. However, public social space in real-world systems is not flat. Individuals organize around particular social coordinates, that reflect characteristics, such as popular religious beliefs, cultures or ideologies. We extend our model by introducing a shape of the social space where sets of "preferred" positions become wells of the social space, such that they act as attracting regions of agents. Depending on the depth and size of the wells, we can distinguish between stronger or weaker attracting parts of the social space. Due to the noise in our model, these wells are not stable sets, but rather metastable sets of the system, meaning that most individuals will spend lots of time within these wells, but rare transitions between the wells are possible. 

We introduce an example of a social landscape given by a double-well potential $W(x,y)=3(x^2-0.25)^2+y^2$, such that social dynamics from ~\ref{eq:ABM} becomes 
\begin{align*}
    dx_k &= \frac{1}{N}\sum^N_{j=1}U(x_k,x_j,\theta_k,\theta_j) dt -\nabla W(x_k)dt+ \sigma_{sp}dW_k^{sp}. \label{eq:DoubleWell}
\end{align*}

\begin{figure}
    \centering
    \includegraphics[width=0.6\textwidth]{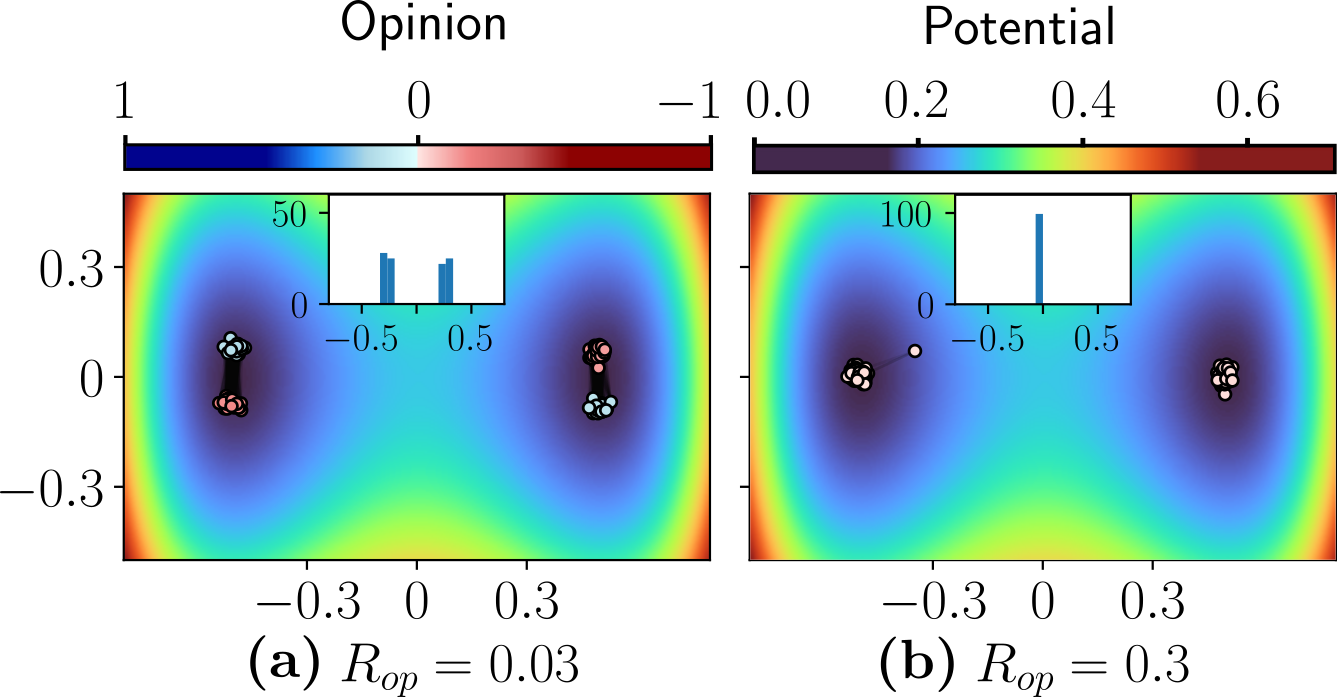}
    \caption{Snapshots of individual simulations at time $T=2.5$ with a double-well potential determining a social space, for different values of opinion interaction radius (a) $R_{op} = 0.03$; and (b) $R_{op} = 0.3$. The colors of agents represent their opinions. The insets show the opinion distribution. Note the different scales of the histograms. The remaining parameters as in Figure~\ref{fig:opinion_radius}.}
    \label{fig:Potential}
\end{figure}

In Figure~\ref{fig:Potential} we plot snapshots of individual simulations at time $T = 2.5$ for different values of $R_{op}$. Shape of the social space drives the agents to the two wells. However, similar effects of radii influence as in Figure~\ref{fig:opinion_radius} can be observed also here. Namely, very small values of $R_{op} = 0.03$, compared to $R_{sp} = 0.15$, lead to polarization within the wells that persist. Large values of $R_{op}$ lead to global consensus of agents that group in both wells. The impact of introducing data-driven social space derived from empirical data will be the topic of future work.

\subsection{Global vs. local model}
Normalization by the total number of agents $N$ can be interpreted as a global scaling of the system \cite{Wang2017noisy,Motsch2014} and is often considered when studying the mean-field limit \cite{GoddardPavliotis2021,DjCKoDj22}. Alternatively, in the non-mean field regime a local scaling can be applied, where the interaction term is normalised by the node degree, i.e. the number of agents that are within the interaction radius \cite{Gomes2023}. Then,  agents who interact with only a small number of other agents are influenced significantly more by such an interaction compared to the case where the agent can interact with a large number of other agents. Local scaling generally leads to a speedup of the dynamics compared to the effects by the global scaling, where the dynamics is slowed down by the, in principle, large number of non-interacting pairs of agents. This type of coupling has been reported to improve synchronization in networks of phase oscillators \cite{motterEnhancingComplexnetworkSynchronization2005} and we observe a similar effect here.

\section{\label{sec:MeanField} The mean-field model}
For very large number of agents, ABM simulations can become extremely expensive \cite{Wang2017noisy,garnier2017,helfmann2021,helfmann2023}, so that parameter estimation and model calibration, as well as validation and sensitivity analysis become unfeasible. In the limit of infinitely many agents, $N\rightarrow\infty$, and when interaction networks are complete (fully-connected), one can derive the mean-field limit of the ABM given by \eqref{eq:ABM}, that results in the following partial differential equation (PDE) for the $\mu_t = \text{Law}(x(t),\theta(t))$
\begin{equation}\label{eq:PDE}
\begin{split}
\partial_t \mu_t(x,\theta) 
=&-\text{div}_x\Big( \mathcal{U}(x,\theta, \mu_t) \mu_t(x,\theta)\Big)\\
&- \text{div}_{\theta}\Big(\mathcal{V}(x,\theta,\mu_t) \mu_t(x,\theta)\Big) \\
&+ \frac{1}{2}\sigma_{sp}^2 \Delta_x \mu_t(x,\theta) + \frac{1}{2}\sigma_{op}^2\Delta_{\theta}\mu_t(x,\theta), 
\end{split}
\end{equation}
where we used the following notation
\begin{align*}\label{short_notation}
    \mathcal{U}(x,\theta, \mu_t) &:= \int_{\R^d \times \R}U(x, y, \theta, \eta)d\mu_t, 
    \\\
    \mathcal{V}(x, \theta, \mu_t) &:= \int_{\R^d \times \R}V(x, y, \theta, \eta) d\mu_t.
\end{align*}
This equation coincides with the standard result \cite{sznitman1991} and is a special case of a result for the original model with the choice of additive noise  \cite{DjCKoDj22}. The mean-field model (PDE model) given by \eqref{eq:PDE} is a reduced model of the ABM that is independent of the number of agents $N$ and as such computationally much more efficient for very large $N$ \cite{helfmann2021}.

\begin{figure}
    \centering
    \includegraphics[width=.6\textwidth]{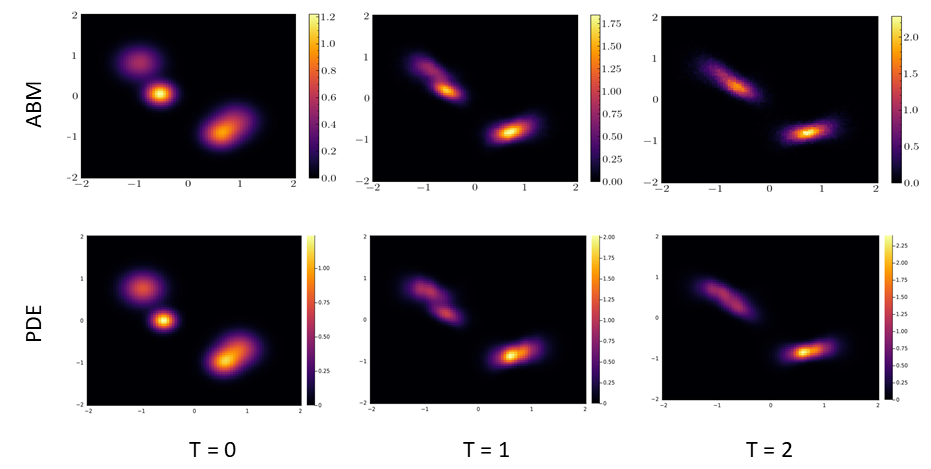}
    \caption{Comparison between the ABM for $N=1000$ agents, averaged over 100 simulations (top row) and the solution of the PDE (bottom row) at three different time points.}
    \label{fig:ABM_PDE}
\end{figure}

Next, we compare the results of the PDE model \eqref{eq:PDE} and the ABM \eqref{eq:ABM_Net} for $N=1000$ agents. Due to the complexity of numerical schemes in higher dimensions, we consider only one dimension in the social space (instead of two) and one opinion dimension. For the mean-filed model, we use the finite difference method. The initial conditions are randomly chosen three clusters with normal distributions in the domain $[-2,2]^2$ and the time interval is $[0,2]$. Parameters are set to $\alpha=\beta =5$, $R_{op}=R_{sp}=0.15$ and $\sigma_{op}^2=\sigma_{sp}^2=0.02$. In the top row of Fig.~\ref{fig:ABM_PDE} we show the results obtained by averaging over $100$ ABM simulations that are visualized using a kernel density estimate (KDE) with Gaussian kernels. We observe a good agreement between the average results of the ABM simulations and the mean-field model.

However, when interaction networks $G_{op}$ and $G_{sp}$ are heterogeneous, deriving generalizations of the PDE \eqref{eq:PDE} become very difficult. In this case graph limits, such as graphons and graphops, could be considered, as they have been used to study simpler systems \cite{kuehn2020network,chiba2016mean,kuehn2022coevolution,KurthsEtAl2023Review}. The case of co-evolving (adaptive) networks remains for future research.

\section{\label{sec:GSSResults}Numerical results on empirical data}
Connecting opinion dynamics models to data and achieving empirical validation remains one of the most important challenges in the field. Parameter estimation is a topic for future research in its own and comes with data availability problems. Ideally, we would have access to panel data containing the opinions of individual agents for several years, which could make precise parameter estimation feasible \cite{SDE_parameter_estimation}. Our main goal in this section is to take the first step in this direction and show that our model is a promising candidate for investigating the feedback loop between social and opinion dynamics in real-world discourse when such a dataset becomes available.

In particular, we will apply our model to depict opinion dynamics based on a real-world dataset obtained from the General Social Survey (GSS) \cite{davern2024gss}. This dataset has been collected through a survey of popular beliefs, attitudes and behaviours in the USA since 1972 and the aim is to keep the survey methodology comparable over time, by adhering to the same sampling methods and question-wording. The GSS changed the survey method in 2021 due to the pandemic from face-to-face interviews to web-based collection. So we will only consider data until 2018, to reduce effects of different sampling strategies.

The GSS covers a variety of topics in a single ballot, that allows us to construct agents' social positions and opinions and compare them with the outcomes of our model. For the social space, we consider a variant of the political spectrum (often called political compass or political map) and extract data on political party affiliation and political views. Data were collected through a rating scale, offering 7 different choices for the party affiliation ranging from Democrat to Republican, and 7 different choices for political view ranging from liberal to conservative. To account for feedback between social and opinion dynamics, we consider opinions on topics that are related to government engagement and political questions. We will keep the social space fixed and consider opinions on two different topics using: 
\begin{itemize}
    \item the "HelpSick" data-set on the question "Should the government cover medical bills?";
    \item the "EqualWealth" data-set on the question "Should the government reduce income differences?". 
\end{itemize}

\begin{figure*}
    \centering
    \includegraphics[width=\textwidth]{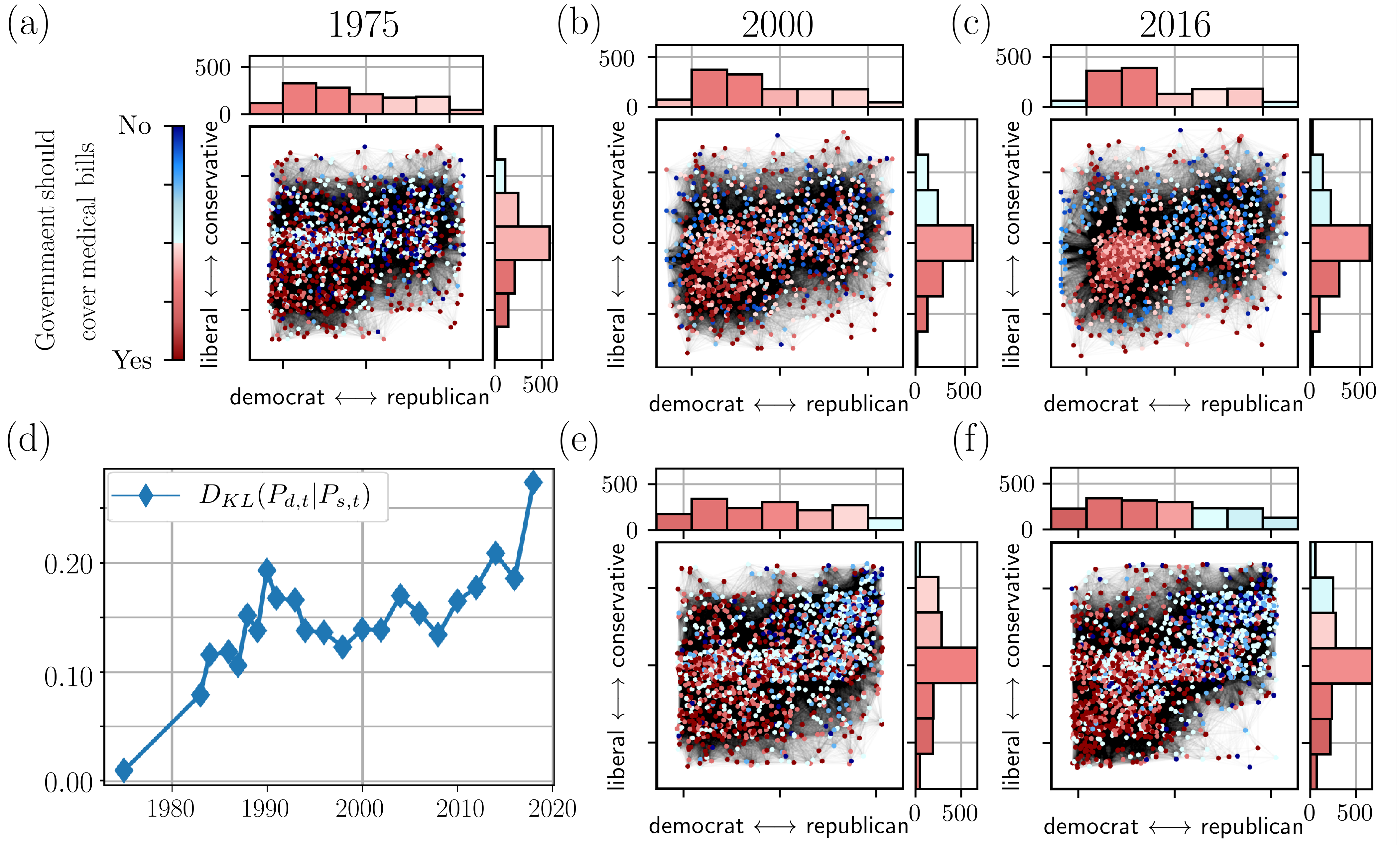}
    \caption{"HelpSick" dataset: Comparison between the simulation results (a,b,c) and the GSS Data (e,f) for the USA election years $2000$ and $2016$. KL divergence for the simulation and data distribution is shown in (d). One realization of the ABM is performed for $N=1355$ agents and the data contains $1687$ and $1774$ responses for the years $2000$ and $2016$ respectively. Other parameters: $R_{op}=0.15 $ $R_{sp}=0.15$ $\alpha=2.7$ $\beta=12.1$ $\sigma_{op}=0.05$, $\sigma_{sp}=0.05$.}
    \label{fig:help_sick}
\end{figure*}

 We start the simulations with the initial conditions generated from the GSS data for each of the questions. More precisely, we fix the number of agents to be the number of valid responses obtained in the first year that the data is available for, i.e. $1975$ for the "HelpSick" and $1978$ for the "EqualWealth" data. To have a consistent scale, we map the answers onto the interval $[-0.25, 0.25]$  for the social space and $[-1, 1]$ for the opinion space. The interaction network, which is determined by the positions of the agents in the social space, plays a crucial role in this model. We construct this network in the initial step by randomly  distributing the agents in the vicinity of the discrete answers from the survey. For details see Appendix \ref{App:Data_Preprosessing}.  Note that the number of valid responses for each question differs over the years and we see large fluctuations in available data. There are different approaches for dealing with this issue, such as imputation methods, but these often produce biases. Since the detailed analysis of this data-set is not at the core of our manuscript, we will leave further data pre-processing for future research. We will nevertheless use it here to demonstrate the potential applicability of our model.

We choose the parameters of the model such that the resulting simulations replicate well the available data. However, the GSS dataset consists of repeated cross-sectional data with long-term information on different samples of individuals each time, meaning that it does not provide the opinion time series of the same individuals. Thus, we compare the GSS distribution $P_{d,t}$ with the distribution coming from a simulation $P_{s,t}$ for each year $t$. Taking into account the data sparsity in some regions of the social space and the continuous values of opinions, we discretize the social space in $3\times 3$ boxes and opinion space in $5$ discrete values. As a result, agents can belong to one of the $45$ possible configurations denoted by $\mathcal{X}$. 
\begin{figure*}
    \centering
    \includegraphics[width=\textwidth]{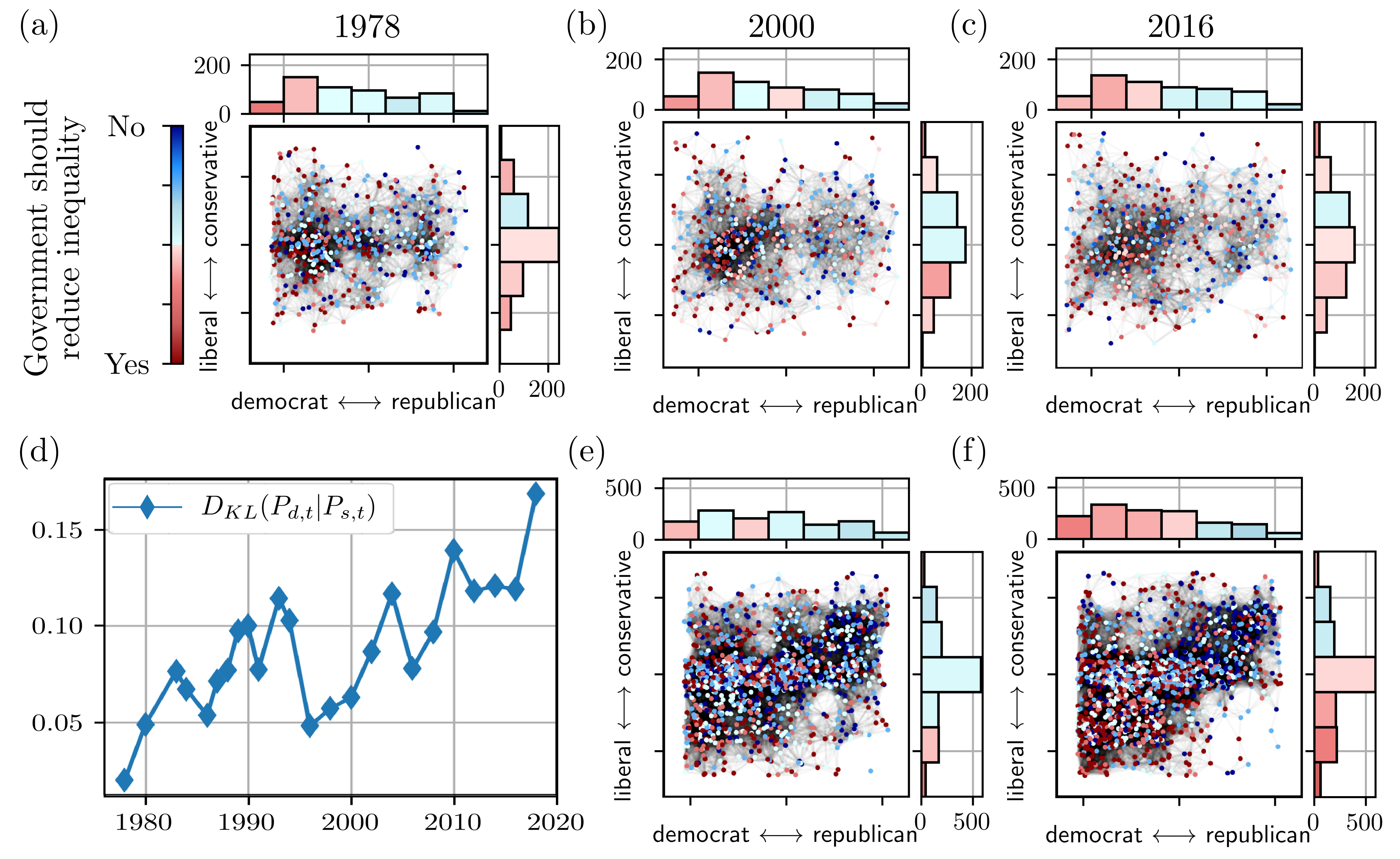}
    \caption{"EqualWealth" dataset: Comparison between the simulation results (a,b,c) and the GSS Data (e,f) for the election years $2000$ and $2016$. KL divergence for the simulation and data distribution is shown in (d). One realization of the ABM is performed for $N=568$ agents and the data contains $1330$ and $1475$ responses for the years $2000$ and $2016$ respectively. Other parameters: $R_{op}=0.08 $ $R_{sp}=0.08$ $\alpha=2.4$ $\beta=15.4$ $\sigma_{op}=0.1$, $\sigma_{sp}=0.1$}.
    \label{fig:equal_wealth}
\end{figure*}
To evaluate the agreement between the data and the simulation results, we use the Kullback–Leibler (KL) divergence
\begin{align}
    D_{KL}(P_{d,t}|P_{s,t}) = \sum_{k\in \mathcal{X}} P_{d,t}(k) \log\left(\frac{P_{d,t}(k)}{P_{s,t}(k)}\right).
\end{align}
and we choose the parameters such that  $\int_{[0,T]} D_{KL}(P_{d,t}|P_{s,t}) \exp(-0.1\cdot t) \,dt$ is minimized. This simple heuristic is chosen such that it favours parameters that give a better fit for times closer to the initialization.
Solving this minimization problem and inferring the suitable parameters is complex and challenging, thus we opt for fixing some of the parameters beforehand. 

In particular, for the "HelpSick" dataset, we set $R_{op}=R_{sp}=0.15$ and $\sigma_{op}=\sigma_{sp}=0.05$, as in the examples above and perform a basic search of the parameter space by randomly sampling ($500$ times) parameters $\alpha, \beta$ from a uniform distribution within the range $[1, 50]$. 
For the "EqualWealth" dataset, we vary the radii as well, so we additionally sample $R_{op}=R_{sp}\sim U([0.03, 0.3])$ and set $\sigma_{op}=\sigma_{sp}=0.1$. The optimal value $R_{op} = R_{sp} = 0.08$ improves the agreement wrt. above considered $R_{op}=R_{sp}=0.15$, producing a different network structure. This indicates that in the case of the "HelpSick" data, networks include more heterogeneous interactions (i.e. with agents that are further apart) than in the case of "EqualWealth". In both cases the strong social influence strength $\beta$ dominates the opinion strength $\alpha$, which has an impact on the network clustering and opinion fragmentation. In Fig.~\ref{fig:help_sick} and Fig.~\ref{fig:equal_wealth} we plot the snapshots of one simulation for the identified parameters and the comparison with the GSS data for the presidential election years $2000$ and $2016$ in the United States. Additionally, we show the total KL-divergence for each year. For clarity, we include the marginal distributions of the social space, where the colour corresponds to the mean opinion of all agents contributing to the specific bin. For snapshots corresponding to all years of the presidential elections between $1978$ and $2016$ see Fig.~\ref{fig:all_election_years} in Appendix\ref{App:Data_Preprosessing}.

For the "HelpSick" data, we observe spatial and opinion clustering within the population from $2000$, both in the data and the simulation, see Fig.~\ref{fig:help_sick} (b,c, e, f). This phenomena is driven by the large value of the social influence (wrt. opinion influence) that guides the system to cluster in a social space, and subsequently to cluster formation in the opinion space. However, the red opinion cluster in the simulation (see (b,c)) is a more prominent compared to the data (see (e,f)). There is an initial increase in the KL divergence, then plateaus between $0.1$  and $0.2$, with a sharp increase between $2016$ and $2018$. Possible causes for this divergence may come from the external factors that have not been considered in the model, such as the introduction of the Affordable Care Act (Obama care). For the "EqualWealth" dataset, we observe more variation in the network structure in Fig.~\ref{fig:equal_wealth}, due to the big difference in the number of agents in the simulation $N=568$ and data, where $N = 1330$ (for year $2000$) and $N = 1475$ (for year $2016$). However, this is not the case for other years (see Fig.~\ref{fig:all_election_years} in Appendix \ref{App:Data_Preprosessing}) and the total KL divergence is smaller than in the "HelpSick" dataset, indicating a better agreement with the data. 

Further interpretation of these results in the context of societal dynamics are out of the scope of this manuscript. Our findings should be seen as the first step towards understanding the complex mechanisms that guide opinion formation process of individuals, influencing and under the influence of their social ties.

\section{\label{sec:Conclusions} Conclusions}
In this paper we study the interplay between the co-evolving opinion and social dynamics in stochastic agent-based models. Our goal is to understand how these dynamics contribute to the opinion formation and complex social behavior observed in real-world systems. We extended state-of-the-art models by introducing a social space where agents move under the influence of both positions and opinions of other agents, driven by the shape of a social landscape, i.e. geography of a social space. Social similarity increases if, agents share similar opinions of the same stance and opinion dissimilarity reinforces social distancing between agents. Opinion dynamics is driven by agents’ opinions and bounded confidence within an interaction radius of agents position in a social space. Unlike existing models that typically rely only on one given social network, we introduce stochastic social dynamics that governs evolution of connections between agents. We distinguish between two co-evolving interaction networks: one governing social dynamics and one for opinion dynamics. Our model highlights the role of these two interaction networks in shaping social systems and opinion formation under their co-evolving dynamics. 

In particular, we show which underlying mechanisms drive the emergence of echo chambers and how these organize within the underlying social space. We observe that small values of opinion interaction radius, lead the system to many fragmented clusters (echo chambers) with local consensus within and diversity between different clusters. Conversely, large radii result in a well-mixed system reaching a consensus state. We quantify the level of polarization in such scenarios by calculating a global network assortativity measure, that considers both the social network and opinion distribution. This measure distinguishes perfectly assortative networks (small opinion interaction radius wrt. social interaction radius) from well mixed systems (much large opinion interaction radius). We perform experiments showing that the number of clusters depend on the model parameters. Our findings reveal a more complex relationship between the number of clusters and opinion interaction radius, compared to what is observed in classical models of this type. Furthermore, we study the system in the limit of infinitely many agents and show the mean-field equation. Numerical simulations demonstrate a good agreement between the ABM and mean-field model. 

Finally, we demonstrate the potential of this model by applying it to empirical survey data on political affiliations and views from the General Social Survey. Our analysis shows the importance of considering co-evolving dynamics when exploring opinion formation on governmental related issues. In particular, our findings indicate for which issues social influence may drive the opinion dynamics or vice versa. 

However, this study is just the first step in understanding the feedback loop between social interactions and opinion formation in real-world discourse. There are several key directions for future research. First, further efforts are needed to bridge the gap between data availability and formal models. Carefully designed studies are needed to build datasets for empirical model validation. We believe that online social media would be an ideal venue for this. Second, extending the model to include data-driven social landscape, influencer agents and multidimensional topic spaces, could provide richer insights into the complex real-world dynamics. Third, going beyond the pairwise interactions and investigating the impact of higher-order interactions, by means of e.g. hypergraphs can offer deeper understanding of social interactions \cite{evolutionary_games_review}. Finally, a problem of deriving mean-field limit equations for the case of two coevovling interaction networks remains open. In this case, network limit could be considered in terms of graphons or graphops, similarly to already studied simpler systems \cite{kuehn2020network,chiba2016mean,kuehn2022coevolution,KurthsEtAl2023Review}. 

\paragraph{Acknowledgments} The authors would like to thank Sebastian Zimper for conducting the numerical experiments on comparison between ABM and PDE model. We are also grateful to Ana Djurdjevac, Philipp Lorenz-Spreen and Christof Schütte for their insightful discussions. This work has been partially funded by the Deutsche Forschungsgemeinschaft (DFG) under Germany’s Excellence Strategy through grant EXC-2046 The Berlin Mathematics Research Center MATH+ (project no. 390685689).

\section*{Data Availability Statement}
All source codes are publicly available 
 \cite{CodeFeedback2024} at \url{https://zenodo.org/records/13270700}. 


\appendix

\section{Further results}\label{app:futher_results}
\begin{figure}
    \centering
    \includegraphics[width=.48\textwidth]{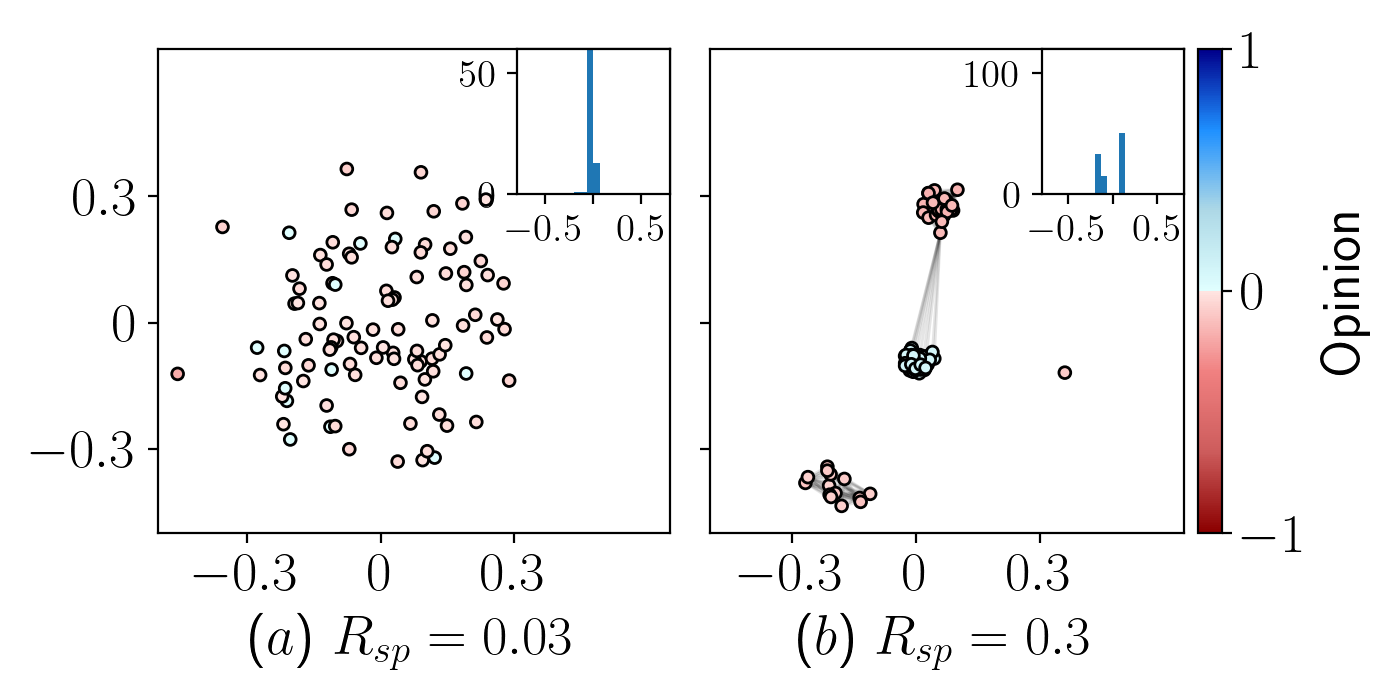}
    \caption{The agents position in the social space for different values of $R_{sp}$ at time $T=2.5$. Lines indicate edges of network $G_{sp}$. The colors represent the opinions of the agents. The insets show the opinion distribution. Note the different scales of the histograms. The remaining parameters: $N=100$, $R_{sp}=0.15$, $\sigma_{op} = \sigma_{sp}=0.05$, $\alpha = 40$ and $\beta= 10$.}
    \label{fig:ChangingR_sp}
\end{figure}

In Fig.~\ref{fig:ChangingR_sp}, we plot the results for varying spatial interaction radius. For small values of $R_{sp}$ compared to $R_{op}$, the system reaches a well mixed state and a global consensus. For very large values of $R_{sp}$, agents group into spatial clusters with local consensus. 

\begin{figure}
    \centering
    \includegraphics[width=0.48\textwidth]{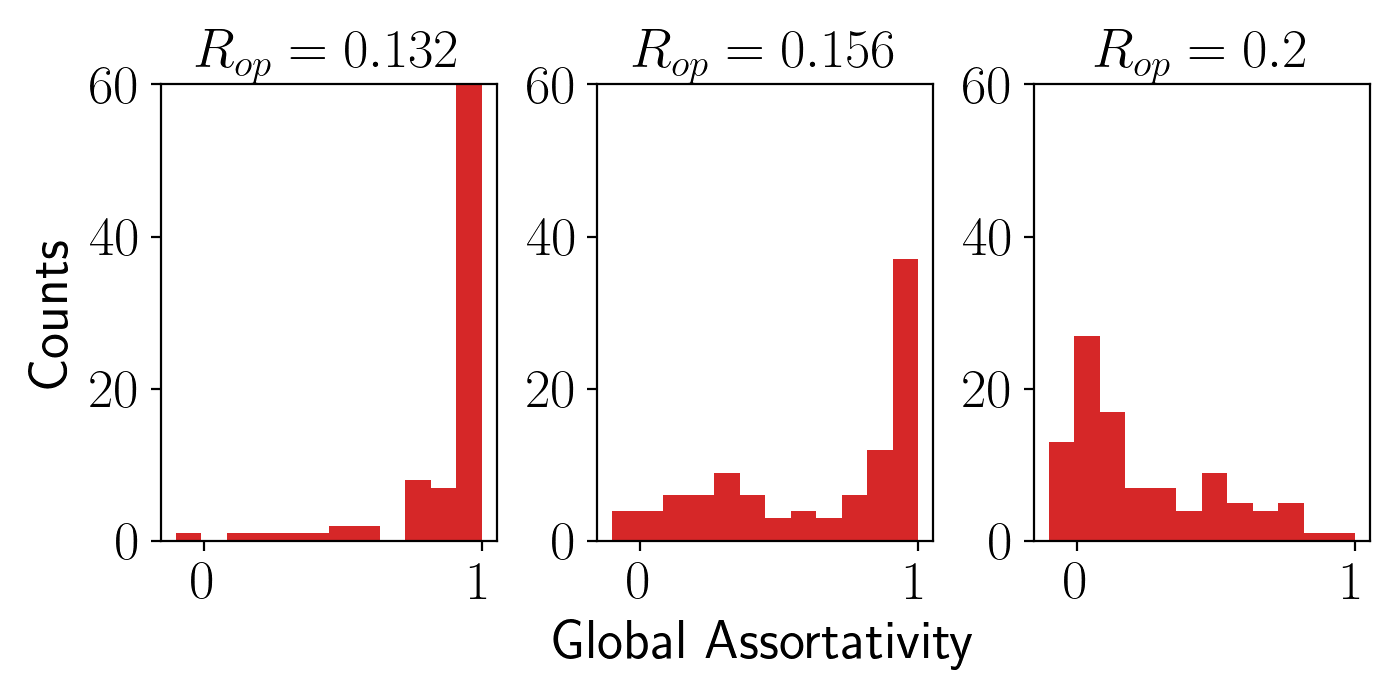}
    \caption{Distributions of the global assortativity for the intermediate regime $R_{op} \approx R_{sp}$. Parameters as in Fig.~\ref{fig:global-assortativity} and generated by averaging over $100$ simulations.}
    \label{fig:bimodality}
\end{figure}

In Fig.~\ref{fig:bimodality}, we plot distributions of the global assortativity for the intermediate regime $R_{op} \approx R_{sp}$ presented in Fig.~\ref{fig:global-assortativity}. When increasing $R_{op}$, the system undergoes a transition form high to low assortativity. During this transition, the  assortativity shows a bimodal distribution, indicating the coexistence of a clustered and a polarized state as well as a consensus state. Parameters as in Fig.~\ref{fig:global-assortativity} and generated by averaging over $100$ simulations.

\section{Data processing\label{App:Data_Preprosessing}}
We start by reducing the GSS dataset to questions about the political views and party alignment of the participants, together with their opinion on coverage of medical bills medical or income differences. For a detailed list of the responses used in our analysis see Table \ref{tab: SocialSpace} and Table\ref{tab:GSS topics}. All data is available on the GSS website and in particular  \href{https://gssdataexplorer.norc.org/variables/846/vshow}{Helpsick data},  \href{https://gssdataexplorer.norc.org/variables/243/vshow}{EqualWealth}, 
 \href{https://gssdataexplorer.norc.org/varia}{PartyID} and \href{https://gssdataexplorer.norc.org/variables/178/vshow}{Political Views}.

\begin{figure}
    \centering
    \includegraphics[width=.48\textwidth]{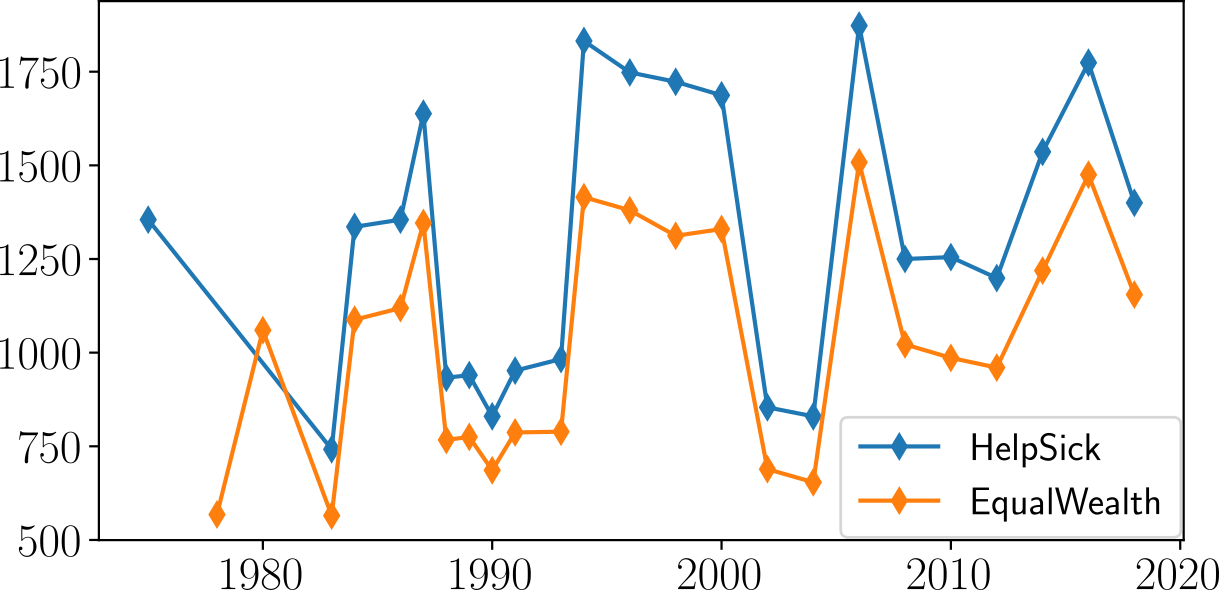}
    \caption{The number of valid responses for the "HelpSick" and "EqualWealth" datasets in different years.}
    \label{fig:participant_numbers}
\end{figure}

 The GSS  employs different ballots and not all participants receive the same set of questions. These answers are marked as "Inapplicable". It also contains responses, that we can't use in this context, since it is not entirely clear how to include parties other than Republicans or Democrats and participants who choose not to answer. We need to assign a concrete opinion and position in the social space to each participant, we exclude all participants whose data contains "Inapplicable",  "No answer", "Do not Know/Cannot Choose", "Skipped on Web", "Independent (neither, no response)" in any of the three relevant questions. For the number of participants with valid responses each year see Fig.~\ref{fig:participant_numbers}.

For each opinion topic, we selected the participants who gave valid answers for the combinations of "PartyID", "PolViews" and the topic in question. The questions that indicate agents' position in the social space are the same for each opinion topic, however, the data we used might not be the same for each topic. Participants who have answered with a valid response to "EqualWealth" might not have given a valid answer to "HelpSick". As a result, the social space data can not be compared between the two examples.

To have a consistent scale, we map the answers onto the interval $[-0.25, 0.25]$  for the social space and $[-1, 1]$ for the topics the agents express their opinions about using a linear Min-Max-Scaler \cite{scikit-learn}.

A drawback of this dataset is, that we only have discrete answers that determine the position of agents in the social space. These positions are however crucially important since they determine the interaction network in the population. For this reason, we define a grid with each combination of the discrete answers in the centre of each grid cell. The agents' position is drawn uniformly at random within their respective grid cells, to obtain more realistic interaction network structures.

Additionally, we enforce reflective boundary conditions for the social space with a 10\% margin of the data spread, because agents leaving this domain can't be matched by any data, but for the parameters presented almost no agents come in contact with the boundary during the simulation.

\begin{figure*}
    \centering
    \includegraphics[width=.75\textwidth]{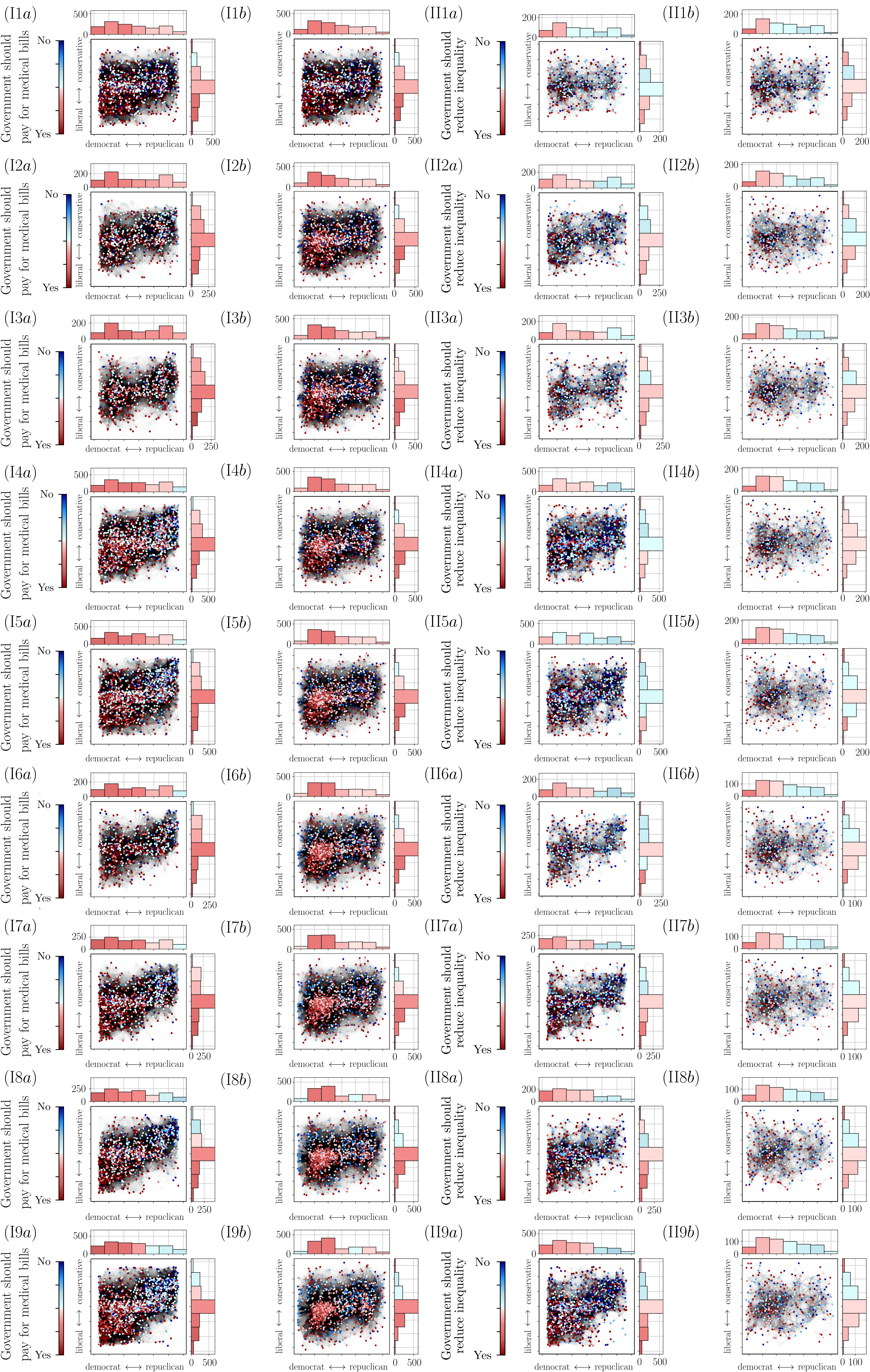}
    \caption{Comparison between the GSS data (a) and the simulation (b) for the election years 1975, 1988, 1990, 1996, 2000, 2004, 2008, 2012, 2016 (1-9) and survey question "HelpSick" (I) and "EqualWealth" (II). Parameters as in Fig.~\ref{fig:help_sick} and Fig.~\ref{fig:equal_wealth}.}
    \label{fig:all_election_years}
\end{figure*}

\begin{table}
\footnotesize
\begin{tabular}{ |p{3.5cm}|>{\centering\arraybackslash}p{1cm}|p{3cm}|>{\centering\arraybackslash}p{1cm}|} 
\hline
\multicolumn{2}{|c|}{\parbox{3.5cm}{\textbf{Question: What is your political party affiliation?} PartyID}} & \multicolumn{2}{|c|}{\parbox{3cm}{\textbf{Question: What are your political views?} PolViews}}\\
\hline
\centering \textbf{Answer} & \textbf{Value} & \centering \textbf{Answer} & \textbf{Value} \\
\hline
Strong democrat & $-0.25$ & Extremely liberal & $-0.25$ \\ 
\cline{1-2}\cline{3-4}
Not very strong democrat & $-0.167$ & Liberal & $-0.167$\\ 
\cline{1-2}\cline{3-4}
Independent, close to democrat & $-0.084$ & Slightly liberal & $-0.084$\\ 
\cline{1-2}\cline{3-4}
Independent (neither) & $0$ & Moderate & $0$\\
\cline{1-2}\cline{3-4}
Independent, close to republican & $0.084$ & Slightly conservative & $0.084$\\
\cline{1-2}\cline{3-4}
Not very strong republican & $0.167$ & Conservative & $0.167$\\
\cline{1-2}\cline{3-4}
Strong republican & $0.25$ & Extremely conservative & $0.25$\\
\hline
\end{tabular}
\caption{Overview of questions in the GSS used for the social space, together with the rescaled values used in the simulations.}
\label{tab: SocialSpace}
\end{table}

\begin{table}
\footnotesize
\begin{tabular}{ |p{3.5cm}|>{\centering\arraybackslash}p{.7cm}|p{3cm}|>{\centering\arraybackslash}p{.7cm}|} 
\hline
\multicolumn{2}{|c|}{\parbox{3.5cm}{\textbf{Question: Should govt help pay for medical care?}\\ HelpSick Data}} & \multicolumn{2}{|c|}{\parbox{3cm}{\textbf{Question: Should govt reduce income differences?}\\EqualWealth Data}}\\
\hline
\centering \textbf{Answer} & \textbf{Value} & \centering \textbf{Answer} & \textbf{Value} \\
\hline
Government should help & $-1$ & 1 Government should reduce differences & $-1$ \\
\cline{1-2}\cline{3-4}
+ &$ -0.5$ & 2 & $-0.67$ \\
\cline{1-2}\cline{3-4}
Agree with both & 0 & 3& $-0.34$\\
\cline{1-2}\cline{3-4}
+ & $0.5$ & 4& $0$\\
\cline{1-2}\cline{3-4}
People should take care of themselves & $1$ & 5& $0.34$\\
\cline{1-2}\cline{3-4}
- & - & 6&$0.67$\\
\cline{1-2}\cline{3-4}
- & - & 7 No government action&$1$\\
\hline
\end{tabular}
\caption{Overview of questions in the GSS used for the opinion distribution, with the rescaled values that are used in the simulations. Answer values: "+" is a possible choice on a scale without specific text; "-" was not an option for the given question and numbers indicate the value on a seven-point scale. }
\label{tab:GSS topics}
\end{table}

\bibliographystyle{plain}
\bibliography{References}
\end{document}